\documentclass[pra,reprint,superscriptaddress,amsmath,amssymb,aps,floatfix]{revtex4-1}

\usepackage{hyperref}
\usepackage{graphicx}
\usepackage{comment}

\usepackage{color}
\usepackage[normalem]{ulem} 	
\definecolor{myblue}{rgb}{0.4, 0.3, 0.7}
\definecolor{purple}{rgb}{0.63,0,1}

\definecolor{dark-green}{rgb}{0,0.4,0.1}
\definecolor{dark-gray}{rgb}{0.4,0.4,0.4}

\definecolor{pink}{rgb}{1,0,0.9}

\newcommand{\Tr}{\operatorname{Tr}}

\usepackage[notext]{stix}
\usepackage{times}

\DeclareSymbolFont{CMAlt}{OMX}{cmex}{m}{n}
\DeclareMathSymbol{\sumop}{\mathop}{CMAlt}{"50}

\usepackage{color}
\hypersetup{%
    unicode=true,
    colorlinks=true,
    linkcolor=blue,
    citecolor=blue,
    urlcolor=blue,
    pdftitle={},
    pdfauthor={I.V. Lukin and A.G. Sotnikov},
    pdfproducer={},
    pdfcreator={}}

\newcommand{\subfigref}[2]{\hyperref[fig:#1]{\ref*{fig:#1}(#2)}}
\newcommand{\subfigrefs}[3]{\hyperref[fig:#1]{\ref*{fig:#1}(#2)--(#3)}}

\begin{document}
\title{Many-body localization in a quantum gas with long-range interactions and linear external potential}

\author{I.V. Lukin}
\affiliation{Karazin Kharkiv National University, Svobody Square 4, 61022 Kharkiv, Ukraine}

\author{Yu.V. Slyusarenko}
\author{A.G. Sotnikov}
\email{a\_sotnikov@kipt.kharkov.ua}
\affiliation{Karazin Kharkiv National University, Svobody Square 4, 61022 Kharkiv, Ukraine}
\affiliation{Akhiezer Institute for Theoretical Physics, NSC KIPT, Akademichna 1, 61108 Kharkiv, Ukraine}

\date{\today}

\begin{abstract}
We study theoretically transitions between the localized and chaotic many-body regimes in one-dimensional quantum lattice systems with long-range couplings between particles and linear external potential. In terms of established criteria characterizing localization, we construct effective phase diagrams for several types of lattice systems with variable amplitude of the external linear tilt and interaction strength. By means of exact diagonalization and time-dependent variational principle numerical approaches we analyze system dynamics after quenches. Our results reveal that the Stark localization without any artificial source of disorder remains stable upon inclusion of long-range interactions.
\end{abstract}

\maketitle

\section{Introduction}
Ergodicity of many-body systems and its breaking is one of the central research areas in modern statistical mechanics.
The basic definition is well understood: the physical system can be called ergodic, if during its time evolution all accessible microstates are visited.
However, the detailed division whether the ergodicity is broken or not, especially, when applied to a large variety of quantum systems is yet to be established~\cite{DAlessio2016}.
A few general examples of ergodicity breaking in these are quantum scars \cite{Bernien2017, Serbyn2021}, Bethe ansatz integrable systems~\cite{QuantumCradle}, lattice gauge theories \cite{LatticeGaugeTheory}, fractons and confinement~\cite{Fractons_confinement}, and Hilbert space fragmentation \cite{Sala2020, PhysRevB.101.174204}.

All these examples of ergodicity breaking and lack of thermalization in closed systems provoke a number of questions on the microscopic characterization of ergodicity in quantum systems. 
In particular, there are interesting connections between thermalization and properties of eigenstates of microscopic Hamiltonian, which are nicely summarized in the form of the eigenstate thermalization hypothesis \cite{SrednickiETH, DeutschETH}, and also with the quasiclassical limit, where these notions overlap with classical chaos \cite{haake1991quantum}.
Rather generally, the notions of quantum chaos and thermalization in closed systems can be used interchangeably, and we thereafter use different criteria of quantum chaoticity, as, in particular, spectral statistics, to detect absence of thermalization. 

As for experimental verification of the mentioned model studies and uncovering new related phenomena,
over the past two decades, the range of accessible quantum many-body systems has been sufficiently extended. 
This progress is largely due to an impressive development of experimental techniques for cooling and loading atoms into optical lattices \cite{Bloch2008RMP, Esslinger2010ARCMP}.
In these artificial systems, many relevant parameters can be controlled and tuned with a high degree of freedom: the external potential (with additional disorder, linear, or any specific), the interaction (both the amplitude and the range), the initial lattice filling, particle statistics, symmetries, etc.
As a natural consequence of this freedom, in particular, the celebrated Anderson localization phenomenon is now viewed as only a member of the wider class of many-body localization (MBL) transitions in interacting systems \cite{Nandkishore2015, Abanin2019RMP}.

From the success of  the Anderson localization, the natural platforms for MBL were initially the systems with disorder.
These platforms were successfully realized in experiments, see, e.g., Ref.~\cite{ExperimentMBL}, where one required artificial disorder produced by quasiperiodic potentials or other means.
Recently, it was shown that analogous systems with only short-range interactions and disorder-free (linear or harmonic) potentials can exhibit localized behavior in a wide range of Hamiltonian parameters, which was named as the Stark (or Bloch) localization  \cite{Nieuwenburg2019, StarkLocalization, Yao2020} and was also observed experimentally \cite{Scherg2021}.

The disorder-free potentials with a linear tilt are common in the field of cold atoms in optical lattices \cite{Raizen1997}.
Typically, the interactions between cold atoms are short-range, however, there are many cases, where these become sufficiently nonlocal, as in gases of atomic isotopes possessing the dipole moment in the ground state~\cite{DipolarGasesOpticalLattice} and atoms in the metastable excited Rydberg states~\cite{Rydberg_atoms_review}. The latter are especially attractive in the context of ergodicity breaking due to quantum scar effects \cite{Bernien2017}.
Furthermore, there is a number of both experimental and theoretical studies on many-body regimes in atomic gases with cavity-mediated interactions \cite{Dogra2016,Landig2016,Sierant2019} (see also the review \cite{CavityQED}). Coupling to the cavity modes in these systems sufficiently extends the effective range of interactions between atoms.

From the theoretical point of view, the combination of the above realizations, namely, the tilted lattice systems, where atoms or quantum spins interact nonlocally, was not studied in detail. 
This motivates us to focus on a wide class of model Hamiltonians, with various types of experimentally available long-range interactions or long-range hopping processes and analyze the fate of many-body localization in these systems.  As we show below, long-range interactions also impact the spectrum in the same regularizing way as an additional disordered or harmonic potential for short-range interacting systems. It turns out that it is sufficient to employ the disorder-free linear external potential with moderate-range interactions between particles to observe and study MBL transitions.

It should be noted that aspects of Stark localization in similar context attracted much interest recently. In particular, there are studies of  the tilted Heisenberg spin chain with the next-nearest couplings \cite{vernek2021robustness}, tilted lattice systems with long-range hopping \cite{PhysRevB.102.085133} and with cavity-mediated interactions \cite{Chanda2021manybody}.

\section{Models and methods}
\subsection{Lattice models}
In this section, we introduce one-dimensional theoretical models in the order of increasing complexity. Starting from the noninteracting limit with only (long-range) hopping and external linear potential, we discuss the localized wave functions and influence of hopping. 
We extend further our description by including power-law interactions, various representations and symmetries of these models. Furthermore, we study a model of localization in all-to-all potentials, which can be realized in cavities \cite{Landig2016}.

\subsubsection{Noninteracting model}
The Hamiltonian of noninteracting lattice model consists of the external linear potential and the hopping term, which describes long-range tunneling processes. Here, we start from an infinite system, i.e., neglect the boundary effects for simplicity,
\begin{equation}\label{eq:Hnonint}
    \hat{H} = - \sum_{j=1}^{m}J_{j}\sum_{k} 
    (\hat{a}^{\dagger}_{k} \hat{a}_{k+j} +{\rm H.c.})  
    + F \sum_{k} k \hat{a}^{\dagger}_{k} \hat{a}_{k},
\end{equation}
where $\hat{a}^{\dagger}_{k}$ and $\hat{a}_{k}$ are bosonic or fermionic creation and annihilation operators on site $k$, respectively. The quantity $F$ characterizes the amplitude of external linear potential and $J_{j}$ are the hopping amplitudes, which depend on the distance~$j$ between the lattice sites. The upper limit~$m$ in the sum denotes the maximal range of hopping. This maximal range can be both finite or infinite in the case of power-law hopping $J_{j} \propto 1/j^\beta$.

The introduced model is quadratic in creation and annihilation operators, thus it is sufficient to solve its one-particle sector. 
Hence, the wave function can be written in the form
\begin{equation}
    |\psi\rangle = \sum_{k} c_{k} \hat{a}^{\dagger}_{k} |0\rangle,
\end{equation}
where $c_{k}$ are the coefficients and $|0\rangle$ is the vacuum state.
We can map the Hilbert space built on the basis states $\hat{a}^{\dagger}_{k}|0\rangle$ onto the Hilbert space of functions on the circle according to the rule \cite{Hartmann_2004}
\begin{equation}
    \hat{a}^{\dagger}_{k} |0\rangle \rightarrow 
    \frac{\exp{\left(i k \phi\right)}}{\sqrt{2 \pi}},
\end{equation}
where $\phi$ is the polar angle.  This results in the mapping
\begin{equation}
    |\psi\rangle 
    \rightarrow 
    \psi(\phi)=\sum_{k} \frac{c_{k} \exp{\left(i k \phi\right)}}{\sqrt{2 \pi}}.
\end{equation}

Within the introduced procedure, it is possible to map operators entering the Hamiltonian~\eqref{eq:Hnonint} to differential operators on the circle. For the linear potential term, the corresponding mapping can be written as follows:
\begin{multline}\label{eq:map1}
    \sum_{m} m a^{\dagger}_{m} a_{m} \sum_{k} c_{k} a^{\dagger}_{k} |0\rangle = \sum_{k} k c_{k} a^{\dagger}_{k} |0\rangle \rightarrow 
    \\ 
    \rightarrow \sum_{k} \frac{ k c_{k} \exp{\left(i k \phi\right)}}{\sqrt{2 \pi}} 
    =
    -i \frac{d}{d\phi} \psi(\phi).
\end{multline}
We see that the external linear potential is mapped to the derivative. 
Finally, let us perform analogous mapping for the hopping terms,
\begin{multline}\label{eq:map2}
    \sum_{m} \hat{a}^{\dagger}_{m} \hat{a}_{m+j} 
    \sum_{k} c_{k} \hat{a}^{\dagger}_{k} |0\rangle 
    = \sum_{k} c_{k+j} \hat{a}^{\dagger}_{k} |0\rangle \rightarrow 
    \\ 
    \rightarrow \sum_{k} \frac{ c_{k+j} \exp{\left(i k \phi\right)}}{\sqrt{2 \pi}} 
    = 
    \exp{\left(-i j \phi \right)} \psi(\phi).
\end{multline}
These terms are mapped to the basis functions multiplied by the range-dependent phase factors. 

By means of the obtained mapping rules \eqref{eq:map1} and \eqref{eq:map2}, the Hamiltonian~\eqref{eq:Hnonint} can be expressed as 
\begin{equation}
    {H} = -2\sum_{j=1}^{m} J_{j} \cos{\left(j\phi\right)} -i F \frac{d}{d\phi}.
\end{equation}
The eigenstates of this Hamiltonian can be determined by solving the first-order differential equation, while the eigenvalues are obtained by the condition that the eigenstates must be periodic functions on the circle.
As a result, we obtain the eigenstates,
\begin{equation}\label{eq:psi_n}
    \psi_{n}(\phi) = \frac{\exp{\left(i n \phi + 2 i \sum_{j=1}^{m} \frac{J_{j} \sin{(j \phi)}}{j F}\right)}}{\sqrt{2 \pi}} ,
\end{equation}
and the eigenvalues
\begin{equation}
    E_{n} = F n, \quad n \in \mathbb{Z}.
\end{equation}

The spectrum of the introduced model is independent of the hopping amplitudes and it is the same as of the Hamiltonian with only a potential term. 
It is natural to suggest that the wave functions in the presence of hopping are continuously connected to the wave functions in the atomic limit (the latter are completely localized on one site). The nonzero hopping processes lead to broadening of the wave functions around that center site with a corresponding exponential decay of the density distribution. 
Below, we also show it more directly by expressing the coefficients~$c_k$ that determine the wave function in the initial basis~$|\psi\rangle$. 

From the form of the wave function~\eqref{eq:psi_n} it is clear that eigenfunctions with $n\neq0$ can be obtained from the eigenstate with $n=0$ simply by translation. 
It can also be deduced from the fact that a commutator of the shift operator with the Hamiltonian results in the shift operator itself. 
Hence, for the models considered below, the shift operator can be viewed as a raising operator: all eigenstates can be obtained by repeated action of the shift operator on a particular eigenstate. For $n=0$, we obtain the following coefficients $c_{k}$ in the initial basis:
\begin{equation}
    c_{k} = \frac{1}{2\pi} \int_{-\pi}^{\pi} \exp{\left(-i k \phi + 2 i \sum_{j=1}^{m} \frac{J_{j} \sin{(j \phi)}}{j F}\right)} d\phi
\end{equation}

In the simplest case of only the nearest-neighbor hopping, these coefficients are determined in terms of the Bessel functions~${\cal J}_k(x)$ as $c_{k} = {\cal J}_{k}\left({2J_{1}}/{F}\right)$. 
In a similar case of hopping only between the next-nearest neighbors ($J_{2}\neq0$, while $J_j=0$ for $j\neq2$), the wave functions vanish for odd $k$, while for even $k$ they are given by $c_{2k} = {\cal J}_{k}\left({J_{2}}/{F}\right)$.
Note that in the case of nearest-neighbor hopping, the localization is generally stable to interactions if $F>2J_{1}$, or if the argument of the Bessel functions is smaller than one.
Note that in this case, the exponential vanishing of the wave functions is clear from the expansion of the Bessel functions into series over small argument $x=2J_1/F$, which gives ${\cal J}_{k}(x) \propto x^{k}(1+O(x^2))$. 
In case of the next-nearest-neighbor hopping, we can conjecture analogously that the localization is stable to interactions if the argument of the Bessel function for the noninteracting wave function is smaller than one, or if $J_{2} < F$. 
More generally, we can conclude that the single particle Stark localization is stable if $2J_{m}<mF$. 
Therefore, to suppress localization, the hopping amplitude between the farther neighbors must be larger than the one between the nearest neighbors, which is rarely the case in cold-atom realizations. 
To check this conjecture, we numerically study localization with nearest and next-nearest hoppings and show that $J_{2}$ starts to influence localization only if it is larger than the amplitude~$J_{1}$. The stability of Stark localization towards long-range hopping was also studied in Ref.~\cite{PhysRevB.102.085133}. 

So far, we have discussed the behavior of noninteracting wave functions, if the argument of the Bessel function is smaller than one ($x<1$). In the noninteracting case, the wave function is localized for all values of couplings $J_m$.
The only change at large $x$ concerns the number of maxima of the wave function and their positions, since with a decrease of $F$ the wave functions exhibit several oscillations before their exponential tails. 
In more general case, the wave functions can be obtained by expansions in Bessel series as discussed in Ref.~\cite{PhysRevA.91.023606}. 

It is also possible to check the case of long-range hoppings given by $J_{m} = J/m^{\beta}$. 
In the numerical exact diagonalization approach for finite systems, we observe that eigenfunctions are localized even at $\beta = 0$ (the case of all-to-all hopping) that contrasts to the Anderson localization, which is stable only for $\beta>1$.

\subsubsection{Models with power-law long-range interactions}
As one can see from the preceding results, the many-body states of the noninteracting model are localized for all values of parameters.
But the localization may not be stable with respect to interactions between particles. 
Now, we introduce the long-range many-body interactions and study the possibility of localization in this system.
The issue of MBL in the presence of long-range interactions is also of conceptual value, since for a long time it was accepted that systems with long-range interactions described by the power-law dependence cannot demonstrate localization features.

\begin{figure}[t]
  \includegraphics[width=\linewidth]{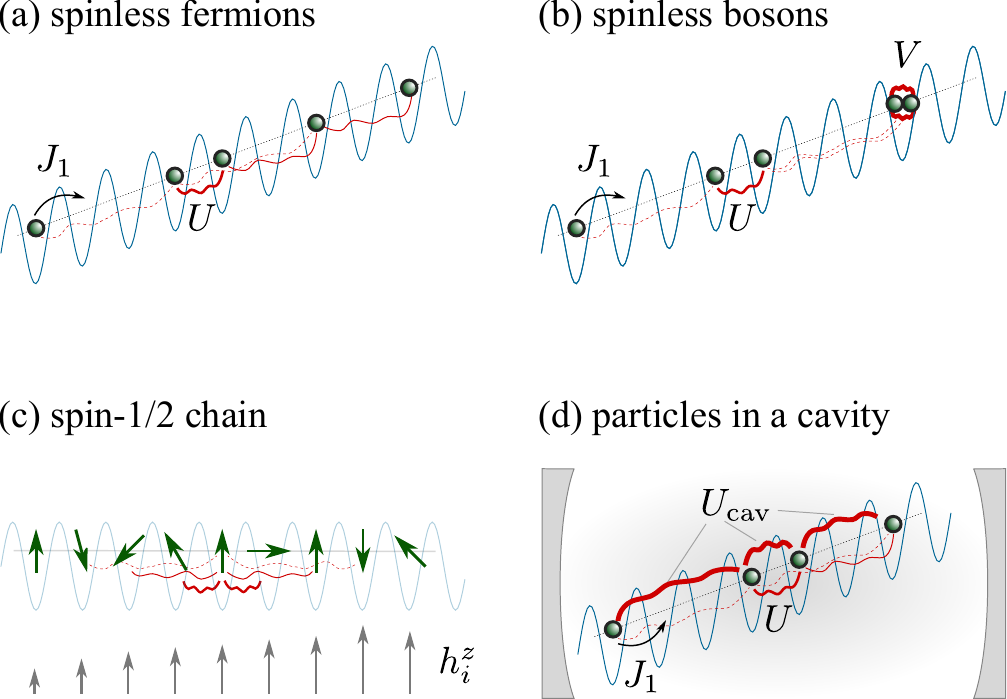}
  \caption{\label{fig:models}%
     Schematic illustration of many-body systems under study and relevant couplings.
     }
\end{figure}

For definiteness, let us introduce the interacting one-dimensional system consisting of spinless fermions on the finite lattice with $L$ sites [see also Fig.~\subfigref{models}{a}].
It is described by the Hamiltonian
\begin{equation}{\label{H_fermionic}}
    \hat{H} 
    = -\sum_{k=1}^{m} J_{k} \sum_{i=1}^{L-k} 
    (\hat{f}^{\dagger}_{i} \hat{f}_{i+k} +{\rm H.c.}) 
    +F\sum_{i=1}^{L} i \hat{n}_{i} + U \sum_{1\leq i < j}^L
    \frac{\hat{n}_{i} \hat{n}_{j}}{|i-j|^{\alpha}},
\end{equation}
where $\hat{f}^{\dagger}_{i}$ and $\hat{f}_{i}$ are the fermionic creation and annihilation operators on site $i$, respectively, and $\hat{n}_{i}=\hat{f}^{\dagger}_{i}\hat{f}^{\phantom{\dagger}}_{i}$ is the corresponding number operator on site $i$. $F$ determines the strength of the external linear potential, as before, $U$ corresponds to the magnitude of interactions between particles, and $\alpha$ is the exponent characterizing the power-law decay of interactions. 
$J_{k}$ are the hopping amplitudes, while $m$ determines the maximal range of hopping as in the noninteracting model. 
Below, we focus on small values of the exponent $\alpha$, in particular, $\alpha \in [0.5,3]$, since at larger values of $\alpha$ the system behaves as the one with short-range interactions. 
The case $\alpha = 3$ is especially relevant, as it can be realized experimentally with dipolar ultracold gases \cite{DipolarGasesOpticalLattice}. 

In case of bosonic system, the Hamiltonian is analogous to Eq.~\eqref{H_fermionic}, except of the additional possibility of the on-site interaction, controlled by the parameter $V$ [see also Fig.~\subfigref{models}{b}]. For the completeness, we specify the explicit form as follows:
\begin{multline}{\label{H_bosonic}}
    \hat{H} = -\sum_{k=1}^{m} J_{k} \sum_{i=1}^{L-k} 
    (\hat{a}^{\dagger}_{i} \hat{a}_{i+k} + {\rm H.c.}) 
    \\ 
    +V\sum_{i=1}^{L} \hat{n}_{i}(\hat{n}_{i}-1)
    +F\sum_{i=1}^{L}  i \hat{n}_{i} + U \sum_{1\leq i < j \leq L} 
    \frac{\hat{n}_{i} \hat{n}_{j}}{|i-j|^{\alpha}},
\end{multline}
where $\hat{a}^{\dagger}_{i}$ and $\hat{a}_{i}$ are the bosonic creation and annihilation operators on site $i$, respectively. 
All other quantities have the same meaning, as in the fermionic case. In numerical calculations, the dimension of the local bosonic Hilbert space has to be restricted to a finite value. We set the maximal number of bosons on the same site equal to three, which is sufficient at moderate values of $V$.

Note that for the purpose of succeeding analysis in the framework of the time-dependent variational principle (TDVP) and the Shrieffer-Wolff transformation, it is necessary to reformulate the fermionic Hamiltonian in the bosonic language. 
For this purpose, we employ the Jordan-Wigner transformation to map the fermionic system onto the spin-1/2 chain. 
In this procedure, the creation and annihilation operators are mapped onto the Jordan-Wigner chains according to the rules: $\hat{f}^{\dagger}_{i} \to \prod_{j=1}^{i-1} (-\hat{\sigma}^{z}_{j}) \hat{S}^{+}_{i}$ 
and $\hat{f}_{i} \to \prod_{j=1}^{i-1} (-\hat{\sigma}^{z}_{j}) \hat{S}^{-}_{i}$, where $\hat{\sigma}^{z}$ is the Pauli matrix with a conventional correspondence to the spin projection operator to the $z$ axis, $\hat{S}^z=\hat{\sigma}^z/2$, while $\hat{S}^{+}=(\hat{\sigma}^x+i\hat{\sigma}^y)/2$ and $\hat{S}^{-}=(\hat{\sigma}^x-i\hat{\sigma}^y)/2$ are the spin-raising and spin-lowering operators, respectively. 
The particle number operator $\hat{n}_i$ is mapped to the local projection operator as $\hat{n}_i \to 1/2 + \hat{S}^{z}_{i}$.
Using these rules, it is possible to map the fermionic Hamiltonian with only the nearest-neighbor hopping ($m=1$) to the following spin-chain Hamiltonian:
\begin{multline}{\label{H_spin}}
    \hat{H} = - J_{1}\sum_{i=1}^{L-1} (\hat{S}^{+}_{i}\hat{S}^{-}_{i+1} + \hat{S}^{-}_{i}\hat{S}^{+}_{i+1})
    + U \sum_{1\leq i < j \leq L} 
    \frac{\hat{S}^{z}_{i} \hat{S}^{z}_{j}}{|i-j|^{\alpha}}
    \\
    + \sum_{i=1}^{L} (F i +w_{i}) \hat{S}^{z}_{i}
     , 
    \quad
    w_{i} = \frac{U}{2}\sum_{j=1, j\neq i}^{L} \frac{1}{|i-j|^{\alpha}}.
\end{multline}
In the given form, this model describes the $XXZ$ spin chain in the external inhomogeneous magnetic field $h_i^z\equiv(Fi+w_i)$, which has additional Ising-type couplings between the spins located farther from each other than nearest neighbors [see also Fig.~\subfigref{models}{c}].
The additional on-site potential term $w_{i}$ becomes constant in the limit of infinite $L$. On the finite lattice, this term is almost constant in the bulk and decreases only at the boundaries.

\subsubsection{Model in a cavity}\label{subsec:cavity}
In addition to the model with power-law long-range interactions, let us also introduce the model of localization in the all-to-all potential. 
This kind of interatomic potential can be realized with a system in a cavity, where presence of the cavity modes, strongly interacting with particles, can induce completely nonlocal interaction patterns. Here, we study a simple model, which captures some basic characteristics of real cavities. In this model [see also Fig.~\subfigref{models}{d}], we introduce an additional term \cite{Dogra2016, Landig2016, Sierant2019, CavityQED, BoseHubbardQuenches}, which is added to the above-specified Hubbard-type Hamiltonians \eqref{H_fermionic} or \eqref{H_bosonic},
\begin{equation}{\label{H_cavity}}
    \hat{H}_{\rm cav} = - \frac{U_{\rm cav}}{L} \left( \sum_{i=1}^{L} (-1)^{i+1} \hat{n}_{i} \right)^{2}.
\end{equation}

This cavity term leads to all-to-all interactions with the same strength between particles irrespectively of the separating distance. 
The interaction amplitude $U_{\rm cav}$ is normalized by the lattice size $L$ to be meaningful in the infinite-lattice limit. 
It is interesting to investigate whether the localization induced by a linear potential is stable to these long-range interactions.

\subsection{Methods}
To distinguish between chaotic and MBL phases, we employ several methods, which are commonly used in the literature on many-body localization. For small system sizes (up to $L = 18$ for spinless fermionic or spin-1/2 systems), it is possible to obtain full spectrum using exact diagonalization. 
Since chaotic and localized systems have different level statistics \cite{Alet2018ManybodyLA}, characteristics of the spectrum can be used as probes of localization.
As the system size grows, exact diagonalization quickly becomes infeasible due to exponential growth of the Hilbert space with the number of the latticce sites~$L$. 
Therefore, below we also employ methods based on the matrix product states to access dynamics of much larger bosonic and fermionic systems after quenches. 
In these simulations, MBL manifests itself as a lack of thermalization of local observables and slow logarithmic growth of the entanglement entropy, similar to observations given, e.g., in Refs.~\cite{Alet2018ManybodyLA, EntanglementGrowth1,EntanglementGrowth2, EntanglementGrowth3, EntanglementGrowth4}. 
For a large linear tilt $F$, we apply the Schrieffer-Wolff transformation to obtain effective Hamiltonians. Within the effective models, we also analyze limiting cases of spectral characteristics and evolution of relevant physical observables.

\subsubsection{Exact diagonalization and level statistics}\label{subsec:ED}
For a small system size and moderate local Hilbert space dimensions (or in a dilute limit, not studied in this work), it is possible to determine full spectrum of the system Hamiltonian in the fixed symmetry sector. The exponential growth of the Hilbert space with a number of lattice sites~$L$ limits these calculations to $L \approx 18$ for spinless fermions, spins or hard-core bosons, or even smaller numbers for bosons with moderate on-site interactions or spinful fermions. 

Chaotic and MBL spectra have different level statistics: Poissonian for MBL phase and Wigner-Dyson for chaotic systems \cite{Alet2018ManybodyLA}. It is connected to the fact that the MBL phase has an extensive set of quasilocal integrals of motion. 
The eigenstates with different eigenvalues of these integrals have uncorrelated energy eigenvalues that leads to the Poisson distribution. However, Hamiltonians of chaotic systems can be represented as random matrices. Eigenvalues of random matrices are distributed according to the Dyson-Wigner ensembles \cite{Mehta2004}. 
Due to the fact that the full spectrum statistics contains an immense amount of information, it is more reasonable to employ a simple quantity, which distinguishes chaotic and MBL systems. 
Such a commonly-used criterion is the gap ratio, usually denoted by $r$. 
To evaluate it, we sort the energy spectrum and calculate the quantity
\begin{equation}{\label{r}}
    {r_i} = \frac{\min(E_{i}-E_{i-1}, E_{i+1} - E_{i})}{\max(E_{i}-E_{i-1}, E_{i+1} - E_{i})}.
\end{equation}
Next, this ratio is averaged over all neighbor triples $\{E_{i-1}, E_{i}, E_{i+1}\}$ in the sorted spectrum, $r=\langle r_i\rangle$.
It is established that $r\approx0.38$ for the MBL systems and $r\approx0.53$ for the chaotic ones \cite{PhysRevLett.110.084101,PhysRevB.75.155111}. 
By analyzing this criterion as a function of system parameters, i.e., by constructing an effective ``phase diagram'', we can determine boundaries between chaotic and localized behavior \cite{PhysRevA.92.041601, PhysRevB.82.174411}.
In this study, we perform the corresponding numerical analysis by means of the \textsc{QuSpin} open-source package \cite{SciPostPhys.2.1.003, 10.21468/SciPostPhys.7.2.020}.

The exact diagonalization (ED) technique can also be employed for studying time dynamics after quenches. 
Within this method, we directly obtain all Hamiltonian eigenstates and project the initial wave function onto them.
Although ED is feasible for small system sizes, its substantial benefit is that the evolution of physical observables can be analyzed on exponentially large timescales (in contrast to the TDVP approach, where the complexity scales linearly with time). 
Access to quantities in this regime allows us to study asymptotic behavior of relevant observables and their fluctuations.

\subsubsection{Ensemble-based analysis}\label{subsec:ensembles}
The physical observables obtained within the ED approach can also be compared with those from the diagonal and microcanonical ensembles \cite{DAlessio2016}. 
This comparison is an explicit test for thermalization or its absence, since for the thermalized system the local observables must stay in agreement with those provided by the microcanonical ensemble. 
The observables from the microcanonical or diagonal ensembles can be evaluated from the available data obtained in the simulation of system dynamics.

For the microcanonical ensemble, we calculate observables in the following way. First, we calculate the expectation value~$E^{(0)}=\langle\psi^{(0)}| \hat{H} |\psi^{(0)} \rangle$ of the Hamiltonian in the initial state~$|\psi^{(0)} \rangle$ before the quench.
Next, we specify the range of energies $\Delta E$ and determine all eigenstates ${\psi_i}$ with the energies $E_i\in[E^{(0)} - \Delta E, E^{(0)} + \Delta E]$.
We choose $\Delta E$ in the way that the number $N_{\rm st}$ of the available eigenstates in the interval is about $N_{\rm st}=50$. 
Finally, we evaluate the expectation values of the operator $\hat{\cal O}$ in the microcanonical ensemble (ME) according to the standard formula, 
\begin{equation}\label{eq:obs_ME}
    \langle \hat{\cal O}\rangle_{\rm ME}
    = {N_{\rm st}}^{-1}\sum_{i} 
    {\langle \psi_{i}|\hat{\cal O}|\psi_{i} \rangle}, 
\end{equation}
where the summation is performed over all eigenstates in the specified energy range. 

The diagonal ensemble describes long-time asymptotics of expectation values \cite{DAlessio2016}. To access it, we calculate all eigenstates $|v_{j}\rangle$ in the symmetry sector (e.g., the block with the fixed total number of particles) of the initial state $|\psi^{(0)}\rangle$.
Next, we calculate the projection coefficients of the initial state onto the eigenstates $\langle v_{j}|\psi^{(0)}\rangle$.
The average values in the diagonal ensemble (DE) are calculated according to the formula
\begin{equation}\label{eq:obs_DE}
    \langle \hat{\cal O}\rangle_{\rm DE}
    =\sum_{j} |\langle v_{j}|\psi^{(0)}\rangle|^{2} \langle v_{j}|\hat{\cal O}|v_{j} \rangle,
\end{equation}
where the summation is performed over all energy eigenstates~$|v_{j}\rangle$. 
Since the long-time asymptotes of physical observables after quench are equal to expectation values in the diagonal ensemble, we compare $\langle \hat{\cal O}\rangle_{\rm DE}$ with $\langle \hat{\cal O}\rangle_{\rm ME}$ to determine whether the system is thermalized. 

As an additional important observable, we also calculate the entanglement entropy. To this end, we consider a state~$|\psi\rangle \equiv|\psi\rangle_{AB}$ and a bipartition of the system $AB$ into two parts: $A$ and $B$ with the respective sizes $L_{A}$ and $L_{B}$.
Then, we can define the entanglement entropy of the subsystem $A$ as the von Neumann entropy of the reduced density matrix $\rho_{A}$ characterizing the subsystem $A$,
\begin{equation}\label{eq:ent_entropy}
    S=-\Tr(\rho_{A}\ln{\rho_{A}}).
\end{equation}
We calculate the reduced density matrix according to the standard formula $\rho_{A} = \Tr_{B}\rho$, where $\rho = |\psi\rangle \langle \psi|$ is the density matrix of the full system under study and the trace is taken over degrees of freedom in the subsystem $B$. 

\subsubsection{Matrix-product state approaches}
As we mentioned above, for large systems the ED approach is not feasible.
However, it is still possible to employ algorithms based on matrix product states (MPS). 
In these methods, it is assumed that the targeted state can be represented as an MPS of a relatively small bond dimension~$D$ (typically, $D\lesssim 100$). There are several classes of such algorithms applicable to MBL systems. 
In particular, the density-matrix renormalization group (DMRG) approach with the corresponding generalization (DMRG-X) can be used to determine eigenstates in the middle of the spectrum of MBL systems \cite{PhysRevLett.116.247204}. 
Unitary matrix-product operator algorithm \cite{PhysRevB.94.041116} finds the full unitary matrix that diagonalizes the Hamiltonian of the localized system.
TDVP \cite{PhysRevB.94.165116} and time evolving block decimation (TEBD) \cite{PhysRevLett.93.040502} can be used to determine dynamics of wave functions after quenches or time dynamics of operators in the Heisenberg picture. 
Below, we restrain ourselves to studying only time dynamics of wave functions. 
Although both TEBD and TDVP approaches can be employed for this purpose, TEBD is restricted to Hamiltonians with short-range interactions. 
As the Hamiltonians of our models contain the long-range terms, TDVP becomes more beneficial for the simulation of quenches. 
In this study, we perform the corresponding tensor-network calculations by means of the \textsc{ITensor} numerical package~\cite{itensor}.

MPS-based approaches are powerful in representing the states with low entanglement entropy. They have a control parameter, the bond dimension $D$, with the maximal entanglement of the representable states, which scales as $\log(D)$. 
One can use unentangled states as initial wave functions, which can be exactly represented as MPS. 
Then, we propagate this state in time within the TDVP approach. 
Naturally, the entanglement entropy increases during the time evolution.
As soon as the entropy reaches approximately the same value as the maximal entanglement entropy for the given bond dimension, results from TDVP become unreliable \footnote{In fact, as results for the system with short-range interactions show, TDVP may become unreliable significantly earlier, see, e.g., the analysis of TDVP convergence in Ref.~\cite{Sierant2021}}. 
For this reason, TDVP is effective only on finite time intervals. However, it can be effectively used for a detection of the MBL regime in which the dynamics significantly slows down. It is much more difficult to unambiguously confirm MBL phase without investigating much longer timescales \cite{Sierant2021}, and there are clear differences between disordered systems and systems with quasiperiodic potentials. The question of how Stark localized systems fit in this scheme needs further investigation, which is beyond the scope of the current study.

For chaotic systems, the entropy increases linearly in time after quenches. Due to this fact, TDVP is applicable on relatively small timescales. In contrast, for MBL systems, the entropy grows logarithmically in time, thus numerical simulations can cover significantly larger time intervals at moderate bond dimensions. 
Moreover, the entropy evolution can be used by itself as a criterion of localization in numerical algorithms. 
Thereafter, the growth of the entropy is used both as an indicator of reliability of the obtained results and as one of representative quantities, which are sensitive to transition between chaotic and localized behavior.

In the subsequent analysis, we use the following quench protocol: we initialize the wave function in the product state, where all even sites of the lattice are filled with one particle and all odd sites are empty (in the case of spin chain, even and odd sites are occupied by spin-up and spin-down particles, respectively); then, the time dynamics of this state for the given model Hamiltonian is studied. In case of fermionic system, we perform the Jordan-Wigner transformation to map the system to the spin chain [see also Eqs.~\eqref{H_fermionic} and \eqref{H_spin}], where one can apply the TDVP approach in a straightforward manner.

While analyzing time evolution of the system, we measure several quantities characterizing the many-body wave function. 
One of them is the above-mentioned entanglement entropy \eqref{eq:ent_entropy}. 
It is also possible to compute expectation values of operators, which can characterize ergodicity breaking. 
Such an experimentally-relevant observable (see, e.g., Ref.~\cite{Scherg2021}), which is especially convenient for our quench protocol, is the even-odd site occupation imbalance~$I$ (or the so-called amplitude of the charge-density wave). It is defined as a difference between the number of particles on even and odd sites of the lattice ($N_{e}$ and $N_{o}$, respectively), normalized by the total number of particles $N$,
\begin{equation}\label{imbalance}
    I(t)
    = \frac{1}{N} \sum_{i=1}^{L} (-1)^{i} 
    \langle\psi(t)|\hat{n}_{i}|\psi(t)\rangle
    = ({N_{e} - N_{o}})/{N}.
\end{equation}

For the above-specified initialization of the wave function, the initial state yields $I(0)=1$ and this is its maximal value, i.e., $|I(t)|\leq1$. 
During the time evolution, this observable typically decreases to a certain constant value and then oscillates around this value with a small amplitude. 
In the chaotic phase, this constant value is close to zero. In contrast, in the MBL phase this value remains relatively large. 
This shows that MBL phase contains some memory of the initial state and its inhomogeneities, which are partly measured by the parameter~$I$. 
Therefore, the asymptotic behavior of imbalance at long times can be used as reliable indicator of localization.

\subsubsection{Schrieffer-Wolff transformation}\label{subsec:SWT}
Let us briefly discuss the case of large external potential in the Hamiltonian.
Note that the corresponding amplitude~$F$ is proportional to the dipole moment of particles or spins. 
The spectrum of the dipole operator entering the Hamiltonian is highly degenerate. Therefore, it seems natural to employ the degenerate perturbation theory based on the Schrieffer-Wolff transformation to effectively describe the system under study. 

The traditional Schrieffer-Wolff transformation (SWT) \cite{sw66} relies on the following procedure: we divide the Hamiltonian into the leading term $\hat{H}_{0}$, which determines the largest energy scale of the full Hamiltonian, and the residual part. 
The latter can be additionally divided into parts $\hat{T}$ and $\hat{V}$ containing operators that do not commute and commute with $\hat{H}_{0}$, respectively.
For definiteness, the spin Hamiltonian~\eqref{H_spin} can be written as
\begin{equation}\label{SW-decomposition}
    \hat{H} = \hat{H}_{0} + \hat{T} +\hat{V},
    \qquad
    \hat{H}_{0} = F\sum_{i=1}^{L}  i \hat{S}^{z}_{i},
\end{equation}
where the part commuting with $\hat{H}_0$ is given by
\begin{equation}
    \hat{V} = \sum_{i=1}^{L} w_{i} \hat{S}^{z}_{i} + U \sum_{1\leq i < j \leq L} 
    \frac{\hat{S}^{z}_{i} \hat{S}^{z}_{j}}{|i-j|^{\alpha}}
\end{equation}
and the noncommuting perturbation has the form
\begin{equation}
    \hat{T} = - J_{1}\sum_{i=1}^{L-1} (\hat{S}^{+}_{i}\hat{S}^{-}_{i+1} + \hat{S}^{-}_{i}\hat{S}^{+}_{i+1}).
\end{equation}

Upon this (or a similar) division, we apply the unitary transformation $\hat{\cal U}$ to the Hamiltonians \eqref{H_fermionic}--\eqref{H_spin}. This unitary transformation is represented in the form $\hat{\cal U} = \exp{\hat{\cal S}}$,  where $\hat{\cal S}$ is anti-hermitian operator. 
The operator $\hat{\cal S}$ is expressed in terms of a series in the expansion parameter (${1}/{F}$ in our case) in the way to cancel terms in the Hamiltonian that do not commute with $\hat{H}_{0}$.
This transformation yields an effective Hamiltonian, which is block-diagonal (up to small higher-order corrections in the expansion parameter), with the size of blocks determined by the degeneracy of $\hat{H}_{0}$. 
For the models with the linear potential we obtain the dipole-conserving Hamiltonians. Note that in the limit of infinite system, the resulting effective Hamiltonian is translationally invariant. In this sense, the systems in linear or quadratic external potentials are close to translational invariance.

Let us now discuss the form of the effective Hamiltonians for the above-specified models. For the spin Hamiltonian~\eqref{H_spin}, we obtain
\begin{multline}\label{H_effective_spin}
    \hat{H}_{\rm eff} = F\sum_{i=1}^{L}  i \hat{S}^{z}_{i} + 
    + \sum_{i=1}^{L} w_{i} \hat{S}^{z}_{i} + U \sum_{1\leq i < j \leq L} 
    \frac{\hat{S}^{z}_{i} \hat{S}^{z}_{j}}{|i-j|^{\alpha}} 
    \\
    + \frac{J_{1}^{2}}{F} \left( \hat{S}_{L}^{z} - \hat{S}_{1}^{z} \right) 
    + \hat{H}_{\rm eff}^{(2)},
\end{multline}
where the explicit form of the second-order terms $\hat{H}_{\rm eff}^{(2)}$ is given in Appendix~\ref{App1} for the sake of compactness.
All terms in the effective model commute with the dipole operator $\sum_{i=1}^{L} i \hat{S}^{z}_{i}$. If only short-range interactions are present, this Hamiltonian is additionally fragmented into noninteracting sectors, as described in Ref.~\cite{Sala2020}.

Either from the effective Hamiltonian \eqref{H_effective_spin} with the inverse Jordan-Wigner transformation, or directly from the Fermi-Hubbard model \eqref{H_fermionic}, the effective Hamiltonian can be written as
\begin{multline}\label{H_effective_fermionic}
    \hat{H}_{\rm eff} 
    = F \sum_{i=1}^{L} i \hat{n}_{i} 
    + U \sum_{1\leq i < j \leq L} 
    \frac{\hat{n}_{i} \hat{n}_{j}}{|i-j|^{\alpha}} 
    \\
    + \frac{J_{1}^{2}}{F} \left( \hat{n}_{L} - \hat{n}_{1} \right)
    + \hat{H}_{\rm eff}^{(2)}.
\end{multline}

Up to quadratic terms in the expansion series over ${1}/{F}$, the generator $\hat{\cal S}$ for the spin model~\eqref{H_spin} has the following form:
\begin{equation}\label{SW_spin}
    \hat{\cal S} 
    = -\frac{J_{1}}{F}  \sum_{i=1}^{L-1} (\hat{S}_{i}^{-} \hat{S}_{i+1}^{+} - \hat{S}_{i}^{+} \hat{S}_{i+1}^{-}) 
    + \hat{\cal S}^{(2)},
\end{equation}
see also Appendix~\ref{App1} for the explicit form of $\hat{\cal S}^{(2)}$.
Note that in the fermionic system, the transformation has a similar form except for the absence of terms with $w_{i}$ in the operator $\hat{\cal S}$. All other terms can be obtained from Eq.~\eqref{SW_spin} by applying the Jordan-Wigner fermionization rules. 

The bosonic model \eqref{H_bosonic} contains an additional on-site interaction term with the coupling $V$, thus the effective Hamiltonian up to linear terms in $1/F$ differs from Eq.~\eqref{H_effective_fermionic} only by the term $V\sum_{i=1}^{L} \hat{n}_{i}(\hat{n}_{i}-1)$. 
At the same time, the explicit forms of the quadratic corrections $\hat{H}_{\rm eff}^{(2)}$ and $\hat{\cal S}^{(2)}$ are substantially different for cases of fermions and bosons; these are given separately in Appendix~\ref{App1}.

\section{Results}\label{sec:results}
\subsection{Spectral characteristics}
\begin{figure}[t]
  \includegraphics[width=\linewidth]{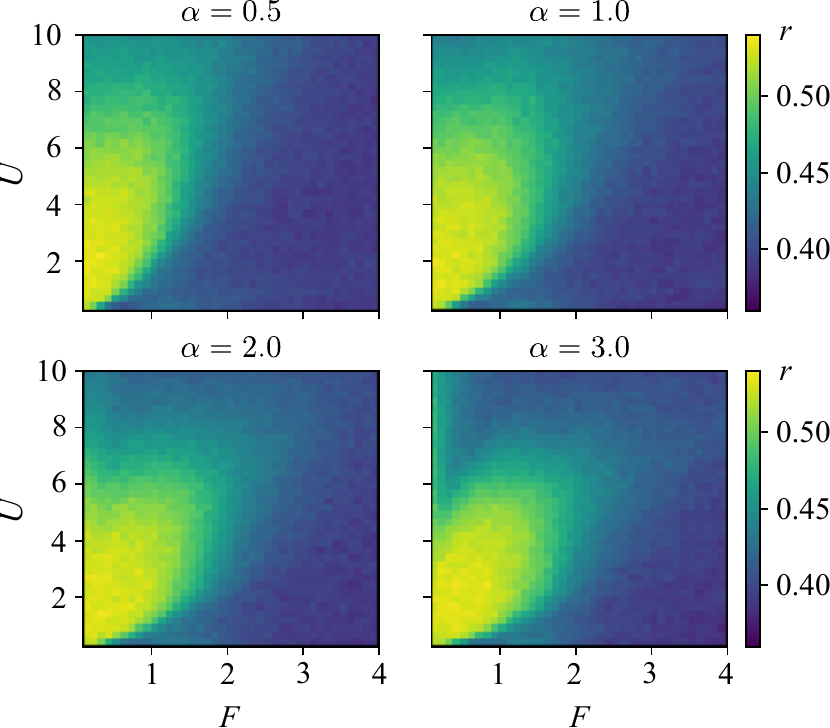}
  \caption{\label{fig:fermionic_spectrum}
    Dependencies of the parameter $r$ on the strength of the external linear tilt $F$ and the interaction strength $U$ at different $\alpha=\{0.5,1,2,3\}$ for the fermionic long-range interacting model~\eqref{H_fermionic} with $L=16$, $N=8$, $J_1=1$, and $m=1$. 
    }
\end{figure}

In this section, we discuss results for the above-introduced ergodicity criterion $r$ [see Eq.~\eqref{r}], which we evaluate by means of the exact diagonalization of various Hamiltonians with long-range deformations of the Hubbard model in the presence of a linear potential (see Subsec.~\ref{subsec:ED}).
We begin our analysis from the Fermi-Hubbard model~\eqref{H_fermionic} with the next-nearest hopping processes ($m=1$).
In Fig.~\ref{fig:fermionic_spectrum} we show the characteristic diagrams of the parameter $r$ in the fermionic system with long-range interactions described by different values of exponents $\alpha$ ranging from $\alpha=0.5$ to $\alpha = 3.0$.
Let us emphasize that the former ($\alpha=0.5$) is far beyond the predicted boundary values of $\alpha$, where localization can occur according to the perturbation theory \cite{Yao2014, Burin2015a, Burin2015b}. Note that there are also ED studies  of the MBL persistence in the presence of similar long-range interactions and aperiodic potentials revealing similar behavior \cite{Nag2019, Prasad2021}.

One of central observations of our study is that the systems with small but nonzero long-range interaction typically remain localized at $F>2$.
This holds in a wide range of the employed parameters $\alpha$ and $U$. 
We attribute it to the fact that long-range interactions completely lift all degeneracies in the spectrum yielding completely regular spectrum statistics with no need for further introduction of the on-site disorder or harmonic potential. 

At large amplitudes of the interaction potential $U$, the systems under study are localized for almost every value of $F$, but this effect is caused rather by conventional Mott-like localization, than by the external linear tilt. 
Since these systems are spinless and do not have internal degrees of freedom, their dynamics is trivial in the strong-coupling limit (in contrast to the effective Heisenberg chains for systems with internal degrees of freedom).

\begin{figure}[t]
  \includegraphics[width=\linewidth]{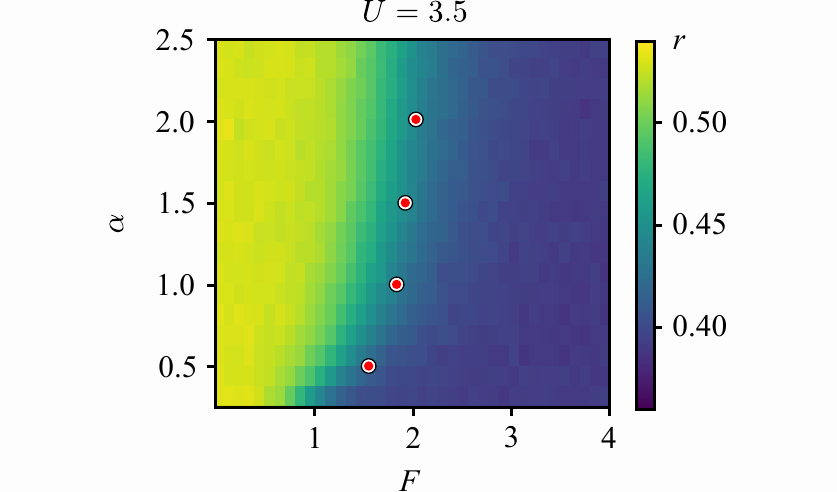}
  \caption{\label{fig:Alpha_dependency}
    Dependencies of the parameter $r$ in the fermionic system on the strength of the external linear tilt $F$ and the exponent $\alpha$ at $L=16$, $N=8$, and $U=3.5$. Red points correspond to the critical amplitude~$F_{c}$ obtained from the finite-size scaling analysis.
    }
\end{figure}

As one can see from Fig.~\ref{fig:fermionic_spectrum}, the chaotic phase is the most pronounced in the interval $U \in (2 ,5)$. To further clarify the influence of the exponent $\alpha$, we fix $U=3.5$ and study the dependence of the gap ratio $r(F, \alpha)$ as shown in Fig.~\ref{fig:Alpha_dependency}. At large values of $\alpha$, the system is effectively short-range and the localization boundary only moderately depends on the exponent~$\alpha$. 
At small $\alpha$, the system becomes additionally localized due to long-range interactions. We use finite-size scaling analysis as described in Appendix~\ref{App2} to extract the critical value $F_{c}$ at different $\alpha$ ~\cite{StarkMobilityEdge, StarkSuperconductingCircuits}. The obtained critical values are indicated by circles in Fig.~\ref{fig:Alpha_dependency}.

Up to this moment, we analyzed stability of localization with respect to introduction of long-range interactions with different power-law dependencies. 
Let us also discuss how long-range hopping can influence MBL. 
To this end, we introduce the next-nearest neighbor (nnn) hopping term with the amplitude $J_{2}$ [see also Eqs.~\eqref{H_fermionic} and \eqref{H_bosonic}] and study its influence on the many-body localization. Note, that the influence of long-range hopping was studied in Ref.~\cite{vernek2021robustness} for the $J_{1}$-$J_{2}$ spin chain in external linear field. Our results agree well with the observations of that study. 

\begin{figure}[t]
  \includegraphics[width=\linewidth]{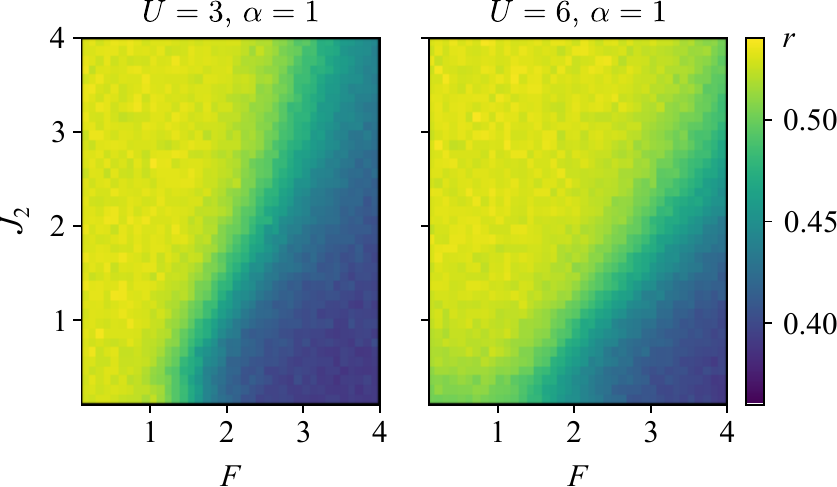}
  \caption{\label{fig:lr_hopping}%
    Dependencies of the parameter $r$ on the strength of the external linear tilt $F$ and the next-nearest-neighbor hopping amplitude $J_{2}$ at $\alpha=1$ and two different $U=\{3,6\}$ for the fermionic long-range interacting model~\eqref{H_fermionic} with $L=16$, $N=8$, $J_1=1$, and $m=2$. 
    }
\end{figure}
Figure~\ref{fig:lr_hopping} shows the dependence of $r$ on the external tilt $F$ and the hopping amplitude $J_{2}$ at two values of long-range interaction strength: $U=3$ and $U=6$. 
In the regime of small $J_{2}$ ($J_{2} \lesssim J_{1}$), the nnn hopping does not impact the localization transition in a visible way. 
Only at $J_{2} > J_{1}$ the transition becomes substantially determined by the amplitude~$J_{2}$. 
In particular, the transition curve exhibits approximately linear dependence of the critical tilt $F_{\rm loc}$ on $J_{2}$, as one can also conclude from the noninteracting model. 
Since in experimental realizations the longer-range hopping terms are usually smaller than the nearest-neighbor terms, the former become largely irrelevant to the issue of stability of Stark localization. We should note that at large interaction strength $U$ the dependence of the gap ratio $r$ on $J_2$ can be more complex, as the system shows localized value of $r$ for all $F$ at $J_{2} =0$. In this case, the nonzero $J_{2}$ can drive the system into the chaotic phase. This behavior is partly shown in Fig.~\ref{fig:lr_hopping} at $U=6$, where the system has intermediate values of $r$ in the limit $J_{2} = 0$.

We have also checked the case of more general power-law hopping $J_{m} = {J_1}/{m^{\beta}}$. The hopping processes parameterized in this way do not destroy localization even at small values of the parameter~$\beta$ ($1< \beta < 2$). 
Note that the stability of MBL was theoretically studied for the lattice model with long-range interactions and the same parametrization of long-range hopping in Ref.~\cite{Nag2019}, but with aperiodic potentials. The given results agree with our observations.

The observed robustness of MBL even upon inclusion of the long-range hopping can be partly explained by arguments based on resonances, which were used to predict breaking of MBL in the case of disordered potentials \cite{Yao2014,Burin2015a,Burin2015b}. 
Resonances are generally present if the difference of energies $|\tilde E_i-\tilde E_{j}|$
between the two eigenstates $|\tilde{\psi}_{i}\rangle$ and $|\tilde{\psi}_{j}\rangle$ of the Hamiltonian without hopping (which includes both many-body interactions and external potential) are smaller than the hopping matrix element between these two respective states. 
In case of an external disorder potential with a randomly distributed amplitude $\epsilon_{n} \in [-W,W]$, there is a nonzero probability that  $|\tilde E_i-\tilde E_{j}|$ is very small, and resonances are present. 
If the number of such resonances diverges with distance between resonating sites, MBL is not stable.
From this, one can derive that for stability of MBL in one-dimensional case with random external potential and local interactions, the condition $\beta > 1$ must be fulfilled.
If more general interactions between the resonances are considered, even more strong restrictions on $\alpha$ and $\beta$ can be obtained. 
For Stark localization, generally, if states $|\psi_{i}\rangle$ and $|\psi_{j}\rangle$ are coupled by a single hopping process between, for example, the sites $m$ and $n$, then the difference between the respective energies $| E_i- E_{j}|$ will be mainly determined by the external tilt, $| E_i- E_{j}| \approx F|m-n|$. 
Resonance will be present only if ${J_1}/{|m-n|^{\beta}} > F|m-n|$, which is generally not the case for large enough $F$ and $|m-n|$ with $\beta > 0$. 
Hence, the natural mechanism of MBL destabilization by long-range hopping is significantly suppressed by the nature of potential, which largely inhibits the possibility of resonances between the states coupled by a single hopping process. 

The above analysis reveals stability of the Stark many-body localization upon inclusion of long-range interactions and long-range hopping processes. The natural question arises on experimentally realistic interaction terms that are able to make Stark MBL unstable or, at least, to shift the localization boundary to the larger values of $F$.
These interactions must contain even the longer-range coupling than in the power-law dependence.
The obvious type of interactions to examine are the cavity-mediated interactions (see Subsec.~\ref{subsec:cavity}), which have already been studied in the context of MBL \cite{Sierant2019,Chanda2021manybody}.
\begin{figure}[t]
  \includegraphics[width=\linewidth]{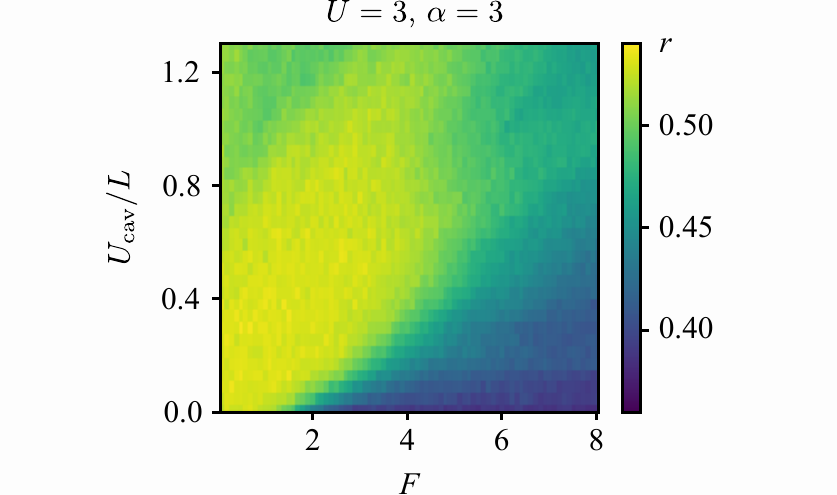}
  \caption{\label{fig:cavity_spectrum}%
    Dependence of the parameter $r$ on the amplitude of the cavity-mediated interaction $U_{\rm cav}$ and the external linear tilt $F$. The fixed parameters are $L=16$, $N=8$, $U=3$, and $\alpha = 3$. 
    }
\end{figure}
In Fig.~\ref{fig:cavity_spectrum} we analyze the ergodicity parameter~$r$ as a function of $F$ and $U_{\rm cav}$, while the amplitudes $U$ and $\alpha$ are kept fixed. 
We observe that the localization boundary shifts to the larger values of $F$ compared to the case of $U_{\rm cav}=0$. 
At larger values of the external tilt $F$ (in particular, $F \gtrsim 5$ for the chosen set of parameters), this system remains localized.

Note that in the limit of large tilt $F$, we also verified the obtained ED results for spectral characteristics by means of the SWT-based calculations [see Subsec.~\ref{subsec:SWT}].
The corresponding analysis confirms, in particular, that the spectral characteristics of the effective Hamiltonians \eqref{H_effective_spin} and \eqref{H_effective_fermionic} are the same as of the full models \eqref{H_fermionic}--\eqref{H_spin} up to corrections proportional to $1/F^{3}$.

\subsection{Dynamics: Imbalance and entropy}
\begin{figure}[t]
  \includegraphics[width=\linewidth]{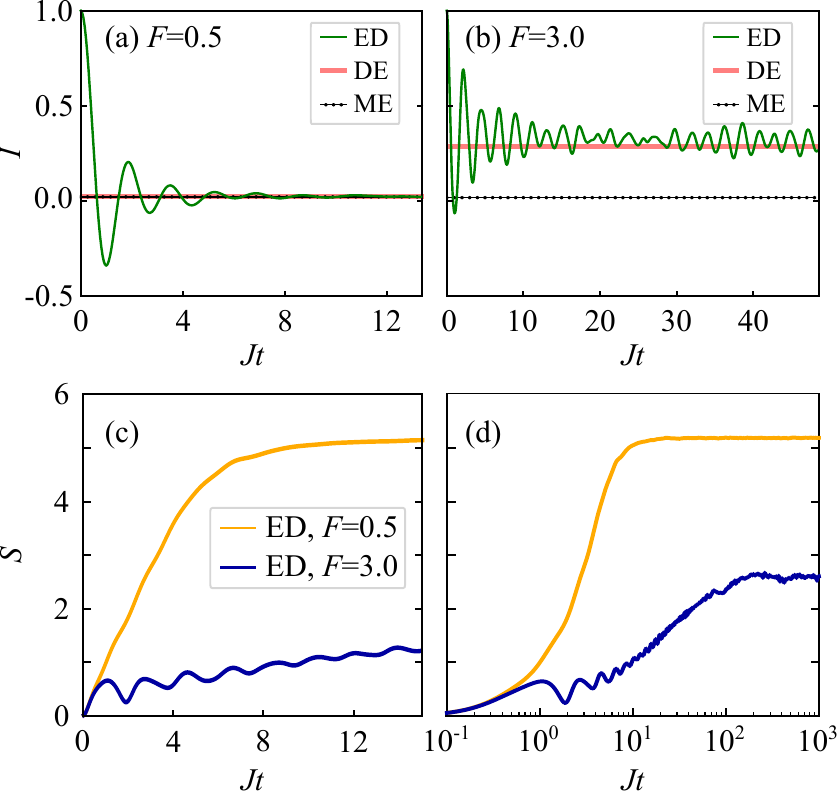}
  \caption{\label{fig:Imbalance_dyn}%
    Time evolution of the imbalance $I$ and entanglement entropy~$S$ after quench in the fermionic model at two different values of the external tilt, $F=0.5$ and $F=3.0$. 
    For sake of visibility, the entropy growth is shown on two timescales: linear (c) and logarithmic (d).
    Other parameters are $L=18$, $U=2$, and $\alpha = 1$. Entanglement entropy is computed for the bipartition of the system on two equal parts.
    }
\end{figure}

The localized and chaotic regimes can be identified by clear signatures in the dynamics of physical observables, such as the particle imbalance $I$ [see Eq.~\eqref{imbalance}].
Characteristic examples of this dynamics are shown in Fig.~\ref{fig:Imbalance_dyn} for the fermionic system in the chaotic ($F=0.5$) and localized ($F=3.0$) regimes. 
Note the difference in timescales used in the corresponding subfigures. 
In the chaotic regime [see Fig.~\subfigref{Imbalance_dyn}{a}], the expectation values in microcanonical and diagonal ensembles are nearly the same and close to zero. 
The initial imbalance relaxes to this expectation value on the timescale of the order of $1/J$ (here and below $\hbar=1$, $J\equiv J_1$, {$N=L/2$}, and $m=1$). After this relaxation, the fluctuations of the imbalance become negligibly small. 
In contrast, in the localized regime with much larger $F$ [see Fig.~\subfigref{Imbalance_dyn}{b}], the diagonal and microcanonical ensembles provide us with different expectation values of the imbalance. 
This observable oscillates around the respective expectation value in the diagonal ensemble for a significantly larger period of time. 

Clear indications of localization can also be observed in the dynamics of the entanglement entropy $S$ [see Eq.~\eqref{eq:ent_entropy}], as we show in Figs.~\subfigref{Imbalance_dyn}{c} and \subfigref{Imbalance_dyn}{d}. 
In the chaotic regime with $F=0.5$, the entanglement entropy grows linearly for a short period of time and then saturates to a constant value. 
The period of the linear growth is approximately the same as a period of relaxation of the imbalance~$I$, see also Fig.~\subfigref{Imbalance_dyn}{a}. 
In contrast, in the localized regime with $F=3$, the entropy~$S$ grows only logarithmically in time and demonstrates characteristic oscillations. 
At much longer times it also saturates, but to a smaller value than in the chaotic regime. A more general discussion of the entanglement growth in long-range interacting localized systems can be found in Ref.~\cite{NonAlgebraicEntanglementGrowth}.

We can use these observations on the behavior of the imbalance~$I(t)$ to study localization transition in more detail. 
Below, we obtain full spectrum of the Hamiltonians under study and calculate the imbalance both from the diagonal and microcanonical ensembles, as discussed in Subsec.~\ref{subsec:ensembles}.
For the purpose of quantifying the observed differences in system dynamics, we introduce an auxiliary ergodicity parameter~$\rho$,
\begin{equation}
    \rho=-\log|\langle I \rangle_{\rm DE}-\langle I \rangle_{\rm ME}|.
\end{equation}

\begin{figure}[t]
  \includegraphics[width=\linewidth]{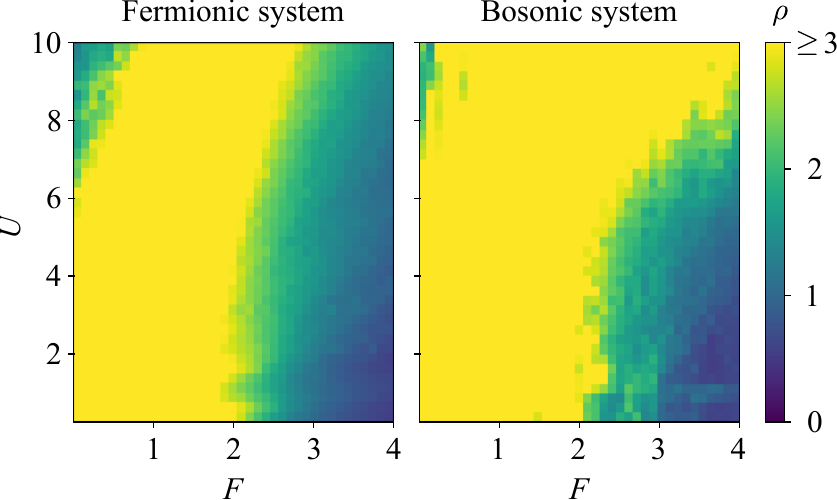}
  \caption{\label{fig:Dynamics_fermions}%
     Dependence of the auxiliary ergodicity parameter $\rho$ on the strength of linear tilt $F$ and strength of long-range interactions $U$ for the fermionic (left, $L=16$, $\alpha = 1$) and bosonic (right, $L=12$, $\alpha=2$, $V=4$) systems.
    }
\end{figure}

In terms of $\rho$, first, we compare predictions given by these two ensembles for the fermionic model~\eqref{H_fermionic} in Fig.~\ref{fig:Dynamics_fermions}.
Here, the boundary between the chaotic and localized behavior at $F\approx2$ and moderate $U$ can be seen much more clearly. 
This boundary starts to shift to larger values of $F$ at higher interaction strength $U$, but it is necessary to note that in this regime both microcanonical and diagonal ensembles predict large values of the final imbalance. 
This shift of the localization boundary to larger interaction strengths is also confirmed by calculations of the entanglement entropy, which exhibits a linear growth to large values even at $F=2.4$ and $U=9$. 

For small values of the interaction strength~$U$, the results from the dynamics and level statistics show some discrepancies (cf. Figs.~\ref{fig:fermionic_spectrum} and \ref{fig:Dynamics_fermions}). 
We further checked the behavior of the entanglement entropy~$S$ in the region of parameter space, where dynamics and level statistics suggest different results. 
The entanglement entropy shows a logarithmic growth to the values typical for chaotic systems, while the imbalance~$I$ fluctuates as in the localized system, but, at the same time, predictions for the mean values $\langle I\rangle_{\rm ME}$ and $\langle I\rangle_{\rm DE}$ agree. 
Furthermore, the microcanonical ensemble predictions become sensitive to the energy range~$\Delta E$ (equivalently, to the number $N_{\rm st}$) used in the definition~\eqref{eq:obs_ME} of the corresponding observables more strongly than in the case of large interaction strength.
In Appendix~\ref{App3} we discuss how the impact of the mentioned discrepancy can be further reduced by analyzing temporal fluctuations of main observables.

Next, for the bosonic model~\eqref{H_bosonic}, we obtain results for the imbalance dynamics. 
Let us note that, according to additional analysis, the results for the spectrum statistics are different at larger values of $V$, as there are states in the spectrum with double or triple occupancies on some sites and these states have energies uncorrelated with other states. 
For bosonic system we restrict ourselves to half-filling to compare with the fermionic case. For larger densities chaotic behavior can survive to higher values of $F$, as effective hopping is enhanced by bosonic statistics.
Therefore, the results from the dynamics become more relevant.
We show the characteristic phase diagram in Fig.~\ref{fig:Dynamics_fermions}. At small and intermediate interaction strength, $\langle I\rangle_{\rm ME}$ and $\langle I\rangle_{\rm DE}$ agree at $F<2$. At larger interactions, as in the fermionic case, microcanonical and diagonal ensembles show similar results only at relatively large tilts $F$. 

\begin{figure}[t]
  \includegraphics[width=\linewidth]{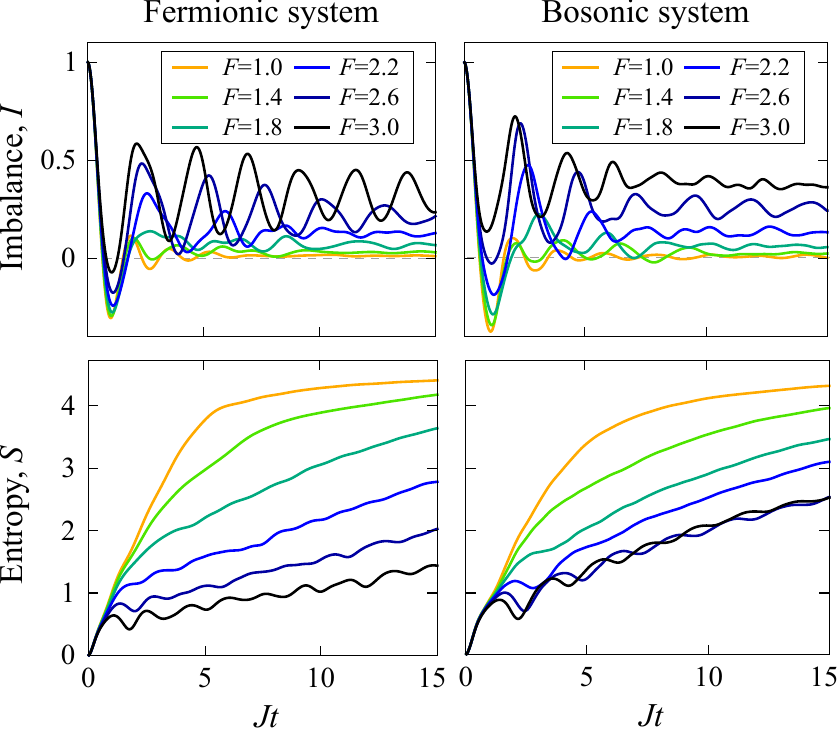}
  \caption{\label{fig:ImbEntTDVP}%
    Time dynamics of the imbalance $I$ and entanglement entropy~$S$ for fermionic (left) and bosonic (right)  systems obtained by TDVP. The parameters are $\alpha = 1$, $U = 3$, $V = 5$ (for bosons), $L = 50$, and $D = 100$. Entanglement entropy is computed for the bipartition of the system on two equal parts.
    }
\end{figure}

\begin{figure}[t]
  \includegraphics[width=\linewidth]{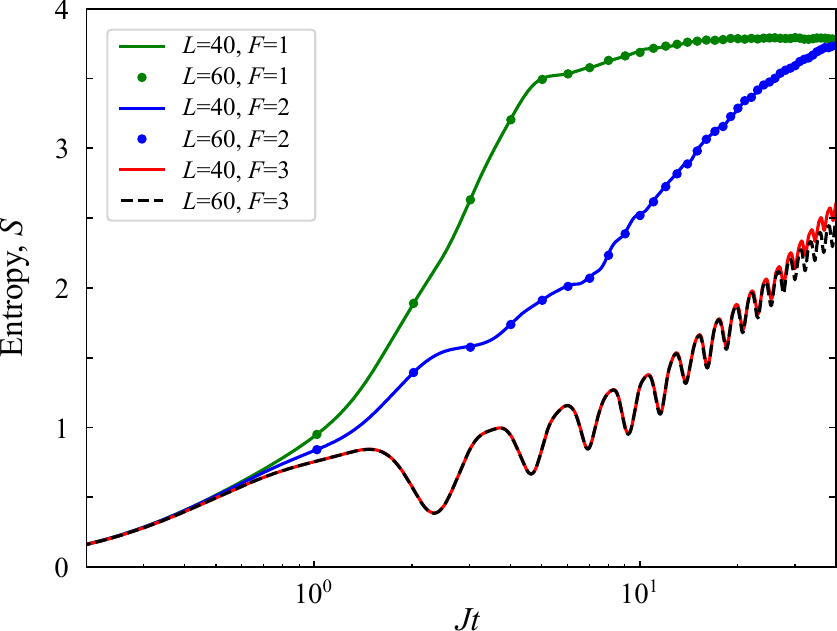}
  \caption{\label{fig:EntTDVPLogScale}%
  Time dynamics of the entanglement entropy on logarithmic timescale in fermionic system for different values of the tilt $F$ and system sizes $L$. The parameters are $\alpha = 1$, $U=3$, and $D=50$. Entanglement entropy is computed for the bipartition of the system on two equal parts.
    }
\end{figure}

For larger systems ($L\geq20$), we employ the TDVP approach to study the imbalance and entanglement entropy behavior. 
In Fig.~\ref{fig:ImbEntTDVP}, we show the results for imbalance dynamics in the bosonic and fermionic systems ($L=50$) at different values of the tilt $F$.  In Fig.~\ref{fig:EntTDVPLogScale}, we additionally analyze the dynamics of entanglement entropy on logarithmic timescale to ensure logarithmic growth of entropy in the localized phase. 

It is clear that the dynamics of imbalance changes for both statistics at $F\approx2$. 
At larger $F$, the imbalance~$I$ exhibits oscillations and relaxes to a nonzero value, while at $F\approx1$ it quickly approaches zero. Behavior is qualitatively the same as was observed in small systems with the exact diagonalization. 
In the TDVP analysis, we employ a relatively small value of the bond dimension ($D=100$), which restricts our calculations at small tilts $F$ to short times, since for larger times the employed MPS approach is not able to accurately represent the amount of entanglement in the wave function. 
This effect can also be seen in Fig.~\ref{fig:ImbEntTDVP}, which shows the growth of the entanglement entropy with time. At small $F$, the entropy reaches the maximum value allowed by the bond dimension~$D$ at several $Jt$. 
This invalidates our results at larger values of $t$,  but also heralds the chaoticity of the system.

\section{Conclusion}\label{sec:conclusion}
We theoretically studied the many-body localization in the case of one-dimensional lattice systems with long-range interaction between particles and linear external potential.
The obtained results reveal that the systems with small but nonzero long-range interaction typically remain localized at moderate and large amplitude of the external linear potential.
These observations hold in a wide range of parameters characterizing long-range interaction potential including the cases of cavity-mediated interactions and long-range tunneling. This makes inclusion of the additional on-site disorder or harmonic potential unnecessary.

In addition to quantitative analysis of spectral characteristics of systems in wide ranges of parameters, we analyzed the dynamical evolution of relevant physical observables: even-odd site occupation imbalance and entanglement entropy.
The dynamics of both quantities clearly indicates differences between the chaotic and localized many-body regimes in lattice systems with the external linear tilt.

Upon calculation of the imbalance within the microcanonical and diagonal ensembles, we introduced an auxiliary (ensemble-based) ergodicity parameter.
For the fermionic systems, we observe qualitative agreement in structures of phase diagrams constructed by means of the ergodicity paramenters from different (spectrum- and ensemble-based) approaches, whereas for the bosonic system, the ensemble-based ergodicity parameter becomes more accurate in certain regimes of the on-site interaction strengths.
Depending on the system size, we applied both ED and TDVP approaches, which agree well in determining localization transitions.
We also confirmed the obtained numerical results in the limit of the applicability of the effective models, where we derived the effective Hamiltonians for the systems under study.

In general, our findings significantly extend the class of systems, where the transitions between the localized and chaotic many-body regimes can be studied in detail by accessing relevant observables in cold-atom experiments \cite{Scherg2021}.
The systems under study are completely disorder free and quasi translationally invariant in the sense that the shift operator commutes with the Hamiltonians up to a constant.
This makes the system identical at different spatial positions and allows one to study it in a kind of thermodynamic limit.
Thus, the approach becomes efficient for the Wegner-flow and Schrieffer-Wolff studies relying on translational invariance \cite{Pekker2017, LongRangeWegnerFlow}.

\begin{acknowledgments}
The authors acknowledge support from 
the National Research Foundation of Ukraine, Grant No.~0120U104963,
the Ministry of Education and Science of Ukraine, Research Grant No.~0120U102252, and the National Academy of Sciences of Ukraine, Project No. 0121U108722.
\end{acknowledgments}

\appendix
\section{Second-order terms in SWT}\label{App1}
Here, let us provide the explicit forms of the higher-order terms ($\propto 1/F^2$) entering the effective Hamiltonians \eqref{H_effective_spin} and \eqref{H_effective_fermionic}, as well as the operator $\hat{\cal S}$
generating the unitary transformation [see Eq.~\eqref{SW_spin}].
The corresponding terms in spin Hamiltonian can be expressed as
\begin{widetext}
\begin{eqnarray}\label{H_effective_spin2}
    \hat{H}_{\rm eff}^{(2)} &=& \frac{J_{1}^{2}}{F^{2}} \biggl[ 
        (w_{2}-w_{1}) \hat{S}_{1}^{z}
    + \sum_{i=1}^{L-1} (w_{i-1} - 2 w_{i} +w_{i+1}) \hat{S}_{i}^{z} + (w_{L-1} - w_{L}) \hat{S}_{L}^{z} \biggr] 
    \nonumber
    \\
    && - \frac{J_{1}^{2}}{F^{2}} \sum_{k=1} ^{L-2} 
    \left[U(k) - U(k+1)\right] 
    \sum_{i=1}^{L-k} \delta _{i} \hat{S}_{i}^{z} \hat{S}_{i+k}^{z} 
    - \frac{2 J_{1}^{2}}{F^{2}} \sum_{k=2} ^{L-1} [U(k) - U(k-1)] \sum_{i=1}^{L-k} \hat{S}_{i}^{z} \hat{S}_{i+k}^{z} 
    \nonumber
    \\
    && - \frac{J_{1}^{2}}{F^{2}} \sum_{k=2}^{L-2} \left[U(k-1) - 2 U(k) + U(k+1)\right] \sum_{i=1}^{L-k-1} \left(\hat{S}_{i}^{+} \hat{S}_{i+1}^{-} \hat{S}_{i+k}^{-} \hat{S}_{i+k+1}^{+}  
    + \rm{H.c.} 
    \right),
\end{eqnarray}
where $U(k) = {U}/{|k|^{\alpha}}$ and $\delta_{i} = 1$, if $i$ is the first or last index in the sum, and $\delta_{i} = 2$ otherwise. The last term can be associated with the dipole-conserving correlated hopping.

The second-order terms in the Fermi-Hubbard Hamiltonian can be written in the form
\begin{eqnarray}\label{H_effective_fermionic2}
    \hat{H}_{\rm eff}^{(2)}
    &=&
    - \frac{J_{1}^{2}}{F^{2}} 
    \sum_{k=1} ^{L-2} [U(k) - U(k+1)]
    \sum_{i=1}^{L-k} \delta _{i} \hat{n}_{i} \hat{n}_{i+k}
    - \frac{2 J_{1}^{2}}{F^{2}} 
    \sum_{k=2} ^{L-1} [U(k) - U(k-1)] 
    \sum_{i=1}^{L-k} \hat{n}_{i} \hat{n}_{i+k}
    \nonumber
    \\
    &&+ \frac{J_{1}^{2}}{F^{2}} 
    \sum_{k=2}^{L-2} \left[U(k-1) - 2 U(k) + U(k+1)\right] 
    \sum_{i=1}^{L-k-1} \left(\hat{f}_{i}^{\dagger} \hat{f}_{i+1} \hat{f}_{i+k} \hat{f}_{i+k+1}^{\dagger}  
    + \rm{H.c.}
    \right).
\end{eqnarray}

The second-order terms entering the generator $\hat{\cal S}$ (valid for both the fermionic and spin models) are
\begin{eqnarray}\label{SW_spin2}
    \hat{\cal S}^{(2)}
    &=&
    - \frac{J_{1}}{F^{2}} \sum_{i=1}^{L-1} (w_{i}-w_{i+1})(\hat{S}_{i}^{-} \hat{S}_{i+1}^{+} - \hat{S}_{i}^{+} \hat{S}_{i+1}^{-}) 
    +\frac{J_{1}U}{F^{2}} 
     \sum_{i=1}^{L-2} \left[
     (\hat{S}_{i}^{-} \hat{S}_{i+1}^{+} - \hat{S}_{i}^{+} \hat{S}_{i+1}^{-})\hat{S}_{i+2}^{z} - \hat{S}_{i}^{z} (\hat{S}_{i+1}^{-} \hat{S}_{i+2}^{+} - \hat{S}_{i+1}^{+} \hat{S}_{i+2}^{-}) 
     \right]
    \nonumber
    \\ 
    &&+ \frac{J_{1}}{F^{2}} \sum_{k=2}^{L-2} \frac{U}{k^{\alpha}} \sum_{i=1}^{L-1-k}(\hat{S}_{i}^{-} \hat{S}_{i+1}^{+} - \hat{S}_{i}^{+} \hat{S}_{i+1}^{-}) \hat{S}_{i+1+k}^{z} 
    - \frac{J_{1}}{F^{2}} \sum_{k=2}^{L-1} \frac{U}{k^{\alpha}} \sum_{i=1}^{L-k}(\hat{S}_{i}^{-} \hat{S}_{i+1}^{+} - \hat{S}_{i}^{+} \hat{S}_{i+1}^{-}) \hat{S}_{i+k}^{z} 
    \nonumber
    \\
    &&+ \frac{J_{1}}{F^{2}} \sum_{k=2}^{L-1} \frac{U}{k^{\alpha}} \sum_{i=1}^{L-k} \hat{S}_{i}^{z} (\hat{S}_{i+k-1}^{-}\hat{S}_{i+k}^{+} - \hat{S}_{i+k-1}^{+} \hat{S}_{i+k}^{-}) 
    - \frac{J_{1}}{F^{2}} \sum_{k=2}^{L-2} \frac{U}{k^{\alpha}} \sum_{i=1}^{L-k-1} \hat{S}_{i}^{z} (\hat{S}_{i+k}^{-}\hat{S}_{i+k+1}^{+} - \hat{S}_{i+k}^{+} \hat{S}_{i+k+1}^{-}).
\end{eqnarray}

Finally, for the bosonic model the second order terms in the Hamiltonian can be expressed as
\begin{eqnarray}\label{H_effective_bosonic2}
    \hat{H}_{\rm eff}^{(2)}
    &=&
    -\frac{J_{1}^{2}}{F^2}
    \left[2U(0)-U(1)\right]
    \sum_{k=1}^{L} \delta_{i}\hat{n}_{i}\left(\hat{n}_{i}-1\right) - \frac{4J_{1}^{2}}{F^2}\left[U(1)-2U(0)\right]
    \sum_{k=1}^{L-1} \hat{n}_{i} \hat{n}_{i+1}
    - \frac{J_{1}^{2}}{F^{2}} \sum_{k=1} ^{L-2} [U(k) - U(k+1)] \sum_{i=1}^{L-k} \delta _{i} \hat{n}_{i} \hat{n}_{i+k}
    \nonumber
    \\
    &&
    - \frac{2 J_{1}^{2}}{F^{2}} 
    \sum_{k=2} ^{L-1} [U(k) - U(k-1)] \sum_{i=1}^{L-k} \hat{n}_{i} \hat{n}_{i+k}
    -
    \frac{J_{1}^{2}}{F^2}\left[2U(0)-2U(1)+U(2)\right]\sum_{i=1}^{L-2} 
    \left(
    \hat{a}_{i}^{\dagger} \hat{a}_{i+1} \hat{a}_{i+1} \hat{a}_{i+2}^{\dagger}  
    + \rm{H.c.} 
    \right)
    \nonumber
    \\
    &&
    - \frac{J_{1}^{2}}{F^{2}} \sum_{k=2}^{L-2} \left[U(k-1) - 2 U(k) + U(k+1)\right]
    \sum_{i=1}^{L-k-1} 
    \left(
    \hat{a}_{i}^{\dagger} \hat{a}_{i+1} 
    \hat{a}_{i+k} \hat{a}_{i+k+1}^{\dagger}
    + \rm{H.c.}
    \right).
\end{eqnarray}

And the second-order terms in the generator $\hat{\cal S}$ for the bosonic system are
\begin{eqnarray}\label{SW_boson2}
    \hat{\cal S}^{(2)}
    &=&
    - \frac{J_{1}}{F^{2}} \left[2U(0)-U(1)\right] \sum_{i=1}^{L-1} (\hat{n}_{i}\hat{a}_{i} \hat{a}_{i+1}^{\dagger} - \hat{a}_{i}^{\dagger} \hat{n}_{i}\hat{a}_{i+1}-\hat{a}_{i} \hat{a}_{i+1}^{\dagger}\hat{n}_{i+1}+\hat{a}_{i}^{\dagger} \hat{n}_{i+1}\hat{a}_{i+1})
    \nonumber
    \\
    &&+
    \frac{J_{1}U(1)}{F^{2}} 
     \sum_{i=1}^{L-2} \left[
     (\hat{a}_{i} \hat{a}_{i+1}^{\dagger} - \hat{a}_{i}^{\dagger} \hat{a}_{i+1})\hat{n}_{i+2} - \hat{n}_{i} (\hat{a}_{i+1} \hat{a}_{i+2}^{\dagger} - \hat{a}_{i+1}^{\dagger} \hat{a}_{i+2}) 
     \right] + \hat{\cal S}_{\rm lr}^{(2)},
\end{eqnarray}
where
\begin{eqnarray*}\label{SW_boson3}
    \hat{\cal S}_{\rm lr}^{(2)}&=& \frac{J_{1}}{F^{2}} \sum_{k=2}^{L-2} U(k) \sum_{i=1}^{L-1-k}(\hat{a}_{i} \hat{a}_{i+1}^{\dagger} - \hat{a}_{i}^{\dagger} \hat{a}_{i+1}) \hat{n}_{i+1+k} 
    - \frac{J_{1}}{F^{2}} \sum_{k=2}^{L-1} U(k) \sum_{i=1}^{L-k}(\hat{a}_{i} \hat{a}_{i+1}^{\dagger} - \hat{a}_{i}^{\dagger} \hat{a}_{i+1}) \hat{n}_{i+k} 
    \nonumber
    \\
    &&+ \frac{J_{1}}{F^{2}} \sum_{k=2}^{L-1} U(k) \sum_{i=1}^{L-k} \hat{n}_{i} (\hat{a}_{i+k-1}\hat{a}_{i+k}^{\dagger} - \hat{a}_{i+k-1}^{\dagger} \hat{a}_{i+k}) 
    - \frac{J_{1}}{F^{2}} \sum_{k=2}^{L-2} U(k) \sum_{i=1}^{L-k-1} \hat{n}_{i} (\hat{a}_{i+k}\hat{a}_{i+k+1}^{\dagger} - \hat{a}_{i+k}^{\dagger} \hat{a}_{i+k+1}).
\end{eqnarray*}

\end{widetext}

\section{Finite-size scaling}\label{App2}
Since finite-size systems are in the main focus of the current study, transitions between the chaotic and localized behavior characterized, e.g., by the changes in the gap ratio $r(F,U,...)$ are completely smooth. 
With an increase of the system size~$L$, the transition features become more sharp. 
One can determine the position of transitions in the parameter space using the finite-size scaling analysis. 
Finite-size scaling is based on the following ansatz for order parameter $A$ (in our context it can be the gap ratio~$r$) as a function of the system size~$L$ and the variable $\sigma$:
\begin{equation}
    A = f\left(L^{{1}/{\nu}}(\sigma - \sigma_{c})\right)
\end{equation}
Here $\sigma_{c}$ is the critical value of the order parameter~$\sigma$ to be determined and $\nu$ is the scaling exponent. 
To determine these values, we employ the numerical package \textsc{pyfssa} \cite{Melchert2009}, which allows for automatic determination of the critical values and critical exponents. 

\begin{figure}[t]
  \includegraphics[width=\linewidth]{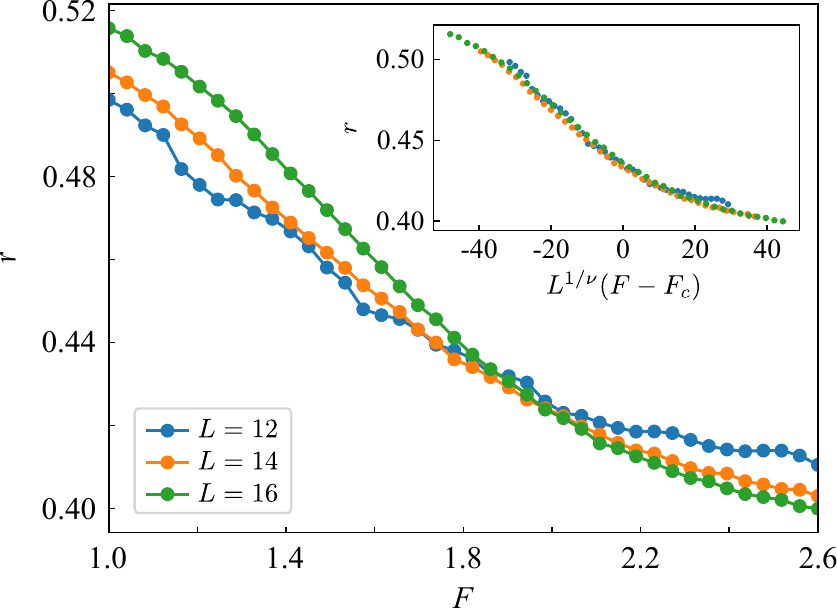}  
  \caption{\label{fig:FSS}%
   Dependencies of gap ratio $r$ on the amplitude~$F$ at different sizes~$L$  of the fermionic lattice system with fixed $U=3.5$ and $\alpha = 1$. Inset shows the same data after rescaling procedure with the calculated critical parameters $\nu=0.68$ and $F_c=1.832$. 
    }
\end{figure}

We take the gap ratio $r$ as the order parameter and determine the critical value $F_c$ of the external linear tilt. To make the behavior of the gap ratio~$r$ completely smooth as a function of the tilt~$F$ and the system size $L=\{12,14,16\}$, we additionally average it over a sufficiently large number of realizations (up to 10$^5$) with a small external disorder potential,
\begin{equation}
    \hat{H}_{d} = \sum_{i=1}^{L} \epsilon_{i} \hat{n}_{i},
\end{equation}
where $\epsilon_{i} \in [-W,W]$ is a random number from the uniform distribution with $W=0.1$. In Fig.~\ref{fig:FSS} we show a characteristic example of the obtained dependencies for $r$ at $\alpha=1$ before and after the finite-size scaling analysis. 

For larger $L$, the influence of the additional disorder potential with the amplitude~$W$ is less pronounced, while at $L=12$ the averaging over disorder realizations does not smooth all fluctuations of $r$, as can be seen in Fig.~\ref{fig:FSS}. We deliberately used small $W$ to minimize its influence on the transition. Additional calculations of dynamical observables and spectral statistics without disorder show similar phase diagram, but extracting transition point and confirming its convergence with system size is more difficult.

\section{Temporal fluctuations in time dynamics}\label{App3}
In the main text we pointed out some discrepancies at small values of the interaction strength~$U$ and small values of the tilt~$F$ between spectral phase diagram in Fig.~\ref{fig:fermionic_spectrum} and results from dynamics obtained by ED in Fig.~\ref{fig:Dynamics_fermions}. We noted that the difference between diagonal and microcanonical ensemble predictions in this parameter range is rather small, while imbalances from both predictions are also small in absolute value. To further quantify the dynamics in this region, one can also study temporal fluctuations of the imbalance as a measure of localization complementary to the difference between diagonal and microcanonical ensembles. 

\begin{figure}[t]
  \includegraphics[width=\linewidth]{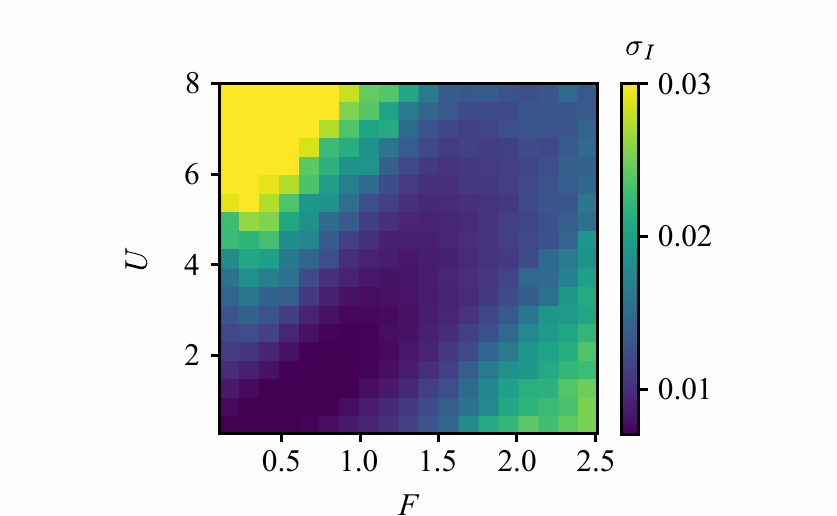} 
  \caption{\label{fig:Fluctuations}%
    Temporal fluctuations $\sigma_{I}$ of the imbalance $I$ in the fermionic system with $L=14$ and $\alpha = 1$. 
    }
\end{figure}

The average temporal fluctuations of the operator $\hat{O}$ are defined as follows \cite{DAlessio2016}:
\begin{eqnarray}
    \sigma_{O} &=& \sqrt{\lim_{t_{0} \to \infty} \frac{1}{t_{0}} \int_{0}^{t_{0}} O(t)^{2} dt - \langle O \rangle_{\rm DE}^{2}} 
    \\ 
    &=& \sum_{m,n, m \neq n} |\langle v_{m}|\psi_{0} \rangle|^{2} |\langle v_{n}|\psi_{0} \rangle|^{2} |\langle v_{m}|\hat{O}|v_{n}\rangle|^{2} ,
\end{eqnarray}
where $O(t)$ is the expectation value of the operator $\hat{O}$ at time $t$ and the vectors $|v_{n}\rangle$ are defined as in Eq.~\eqref{eq:obs_DE}. 
In Fig.~\ref{fig:Fluctuations} we show temporal fluctuations of the imbalance $I$. Fluctuations generally grow in the localized phase (including the region of small $U$ and $F$) and are very small in the chaotic phase. 

\bibliography{mbl_long_range}

\end{document}